\pdfoutput=1

\documentclass[10pt,letter,twocolumn]{article}
\usepackage{f1000_styles}

\usepackage[super,comma,sort]{natbib}
\usepackage[colorlinks=true,urlcolor=blue]{hyperref}

\usepackage{amsmath}

\newcount\BoxNum \BoxNum 1\relax
\makeatletter
\newcommand*{\boxlabel}[1]{%
  \protected@write \@auxout {}{\string \newlabel {box:#1}{{\the\BoxNum}}{}}%
  \advance\BoxNum 1\relax}
\makeatother

\newcommand*{\boldrule}{\hrule height 1.2pt}
\newcommand*{\noterule}{\medskip\boldrule\medskip}	
\newcommand{\noterulenote}[1]{\bigskip\boldrule\nobreak\medskip\nobreak%
	{\vbox{\bf{\noindent #1}}}\nobreak%
	\medskip\nobreak\boldrule\medskip}

\def\citeyear{\citep}
\def\autocite{\citep}
\def\textcite{\citet}

\begin{document}

\title{Puzzles in modern biology. II. Language, cancer and the recursive processes of evolutionary innovation}
\author[1]{Steven A.~Frank}
\affil[1]{Department of Ecology and Evolutionary Biology, University of California, Irvine, CA 92697--2525 USA, safrank@uci.edu}

\maketitle
\thispagestyle{fancy}

\parskip=4pt

\begin{abstract}

Human language emerged abruptly. Diverse body forms evolved suddenly. Seed-bearing plants spread rapidly. How do complex evolutionary innovations arise so quickly? Resolving alternative claims remains difficult. The great events of the past happened a long time ago. Cancer provides a model to study evolutionary innovation. A tumor must evolve many novel traits to become an aggressive cancer. I use what we know or could study about cancer to describe the key processes of innovation. In general, evolutionary systems form a hierarchy of recursive processes. Those recursive processes determine the rates at which innovations are generated, spread and transmitted. I relate the recursive processes to abrupt evolutionary innovation.

\end{abstract}


\vskip0.5in
\noterule
Preprint of published version: Frank, S. A. 2016. Puzzles in modern biology. II. Language, cancer and the recursive processes of evolutionary innovation. F1000Research 5:2089, \href{http://dx.doi.org/10.12688/f1000research.9568.1}{doi:10.12688/f1000research.9568.1}. Published under a Creative Commons \href{https://creativecommons.org/licenses/by/4.0/}{CC BY 4.0} license.
\smallskip
\noterule

\clearpage

\subsection*{Introduction}

Major evolutionary innovations arise abruptly. Novel body forms appeared suddenly, \textit{the Cambrian explosion}\autocite{smith13causes}. Seed-bearing plants spread across the earth almost instantaneously, \textit{Darwin's abominable mystery}\autocite{friedman09the-meaning}. Humans spoke, made symbolic art and dominated the world.

A review\autocite{tattersall16at-the-birth} of \textit{Why Only Us: Language and Evolution}\autocite{berwick15why-only} emphasizes the recurring controversy over evolutionary innovation:\smallskip
\begin{quote}
Today, opinion on the matter of language origins is still deeply divided. On the one hand, there are those who feel that language is so complex, and so deeply ingrained in the human condition, that it must have evolved slowly over immense periods of time. $\dots$ On the other, there are those like Berwick and Chomsky who believe that humans acquired language quite recently, in an abrupt event.
\end{quote}
\vskip-4pt
The argument for slow evolution appeals to intuition. Such complexity cannot evolve suddenly. Evolution is an intrinsically slow process. 

Against the intuitive argument for the slow evolution of language, the evidence suggests that:\autocite{tattersall16at-the-birth}\smallskip
\begin{quote}
Clearly, something revolutionary had happened to our species $\dots$ All of a sudden, humans were manipulating information about the world in an entirely unprecedented way, and the signal in the archaeological record shifted from being one of long-term stability to one of constant change $\dots$ by fifty years ago we were already standing on the moon. $\dots$ So we need an explanation for the abrupt emergence of language $\dots$
\end{quote}

My theme concerns the general understanding of evolutionary process. How surprising is abrupt evolutionary innovation? How do we understand what `abrupt' means? To answer those questions, we must understand the nature of time in relation to generative process.

\subsection*{The abruptness of recursive growth}

Think about cancer. A tumor evolves by accumulating changes\autocite{weinberg07the-biology}. The initial changes may arise before one notices any sign of tumor or disease. Eventually, the tumor acquires novel traits that give it an uncontrolled growth advantage. Overwhelming disease soon follows.

Without modern technology, one sees tumors as arising abruptly. That suddenness comes from the growth rate of tumors, shaped by the history of evolutionary innovations. Synergism between growth and innovation sets the tempo at which we perceive novelty.

Growth by itself has a natural tempo that causes things to appear suddenly. In uncontrolled growth, an initial input size is multiplied by a growth factor, producing a bigger output size. The output then becomes the input for another round of recursive growth. 

A recursive doubling in size produces a series of $1,2,4,8,16,32,\dots$, with a size of $2^n$ at the $n$th time step. A tumor typically must have billions of cells before it is noticed. To grow from one cell to a noticeable size of 30 billion cells, a recursively growing tumor must pass through 35 doubling periods. 

After just 5 more rounds of doubling growth, the tumor will be 32 times larger than the size at first detection. The time is short from being noticed to being overwhelmingly dominant. 

Seemingly abrupt appearance is a property of recursive growth. Put another way, the natural timescale of growth is explosive, whereas the natural timescale of our perception seems to be relatively steady. The perception of appearance by growth tends to be abrupt.

\subsection*{Evolutionary innovation}

I invoked uncontrolled growth. But where does such growth come from? What is the nature of innovation that increases growth?

We may never know the answer for language. At present, we do not know the answer for tumors, even though tumors happen all the time right under our own skin. But perhaps the puzzle of evolutionary innovation in tumors will be solved one day\autocite{frank07dynamics}. 

Deeper understanding of evolutionary innovation in tumors may provide insight into what it takes, more generally, for the origin and spread of seed-bearing plants, of new body forms and of language. So I continue to discuss tumors. The abruptness of cancer is a model of evolutionary innovation.

We know that an aggressive tumor has acquired many evolutionary changes when compared to its normal ancestral tissue. Did most of those cancerous changes happen abruptly around the transition to perceptible aggressiveness? Or did many evolutionary changes accumulate slowly, over a long period, starting well before noticeable cancer? 

We do not know exactly. But we can say what the likely processes are for evolutionary change in cancer, what the timescales are for those processes, and how the different processes interact. We can draft a rough solution to the puzzle of evolutionary innovation in cancer.

I step through the key evolutionary processes and their consequences for the timescale of cancer. At first, the puzzle of cancer may seem rather distant from the puzzle of language. However, consider two questions.  

Is language an example of the known processes of evolutionary innovation? Or does the puzzle of language require a unique solution? We can discuss those questions in a more informed way after briefly considering cancer. 

A successful tumor gains the ability to break through tissue barriers, survive in novel environments, escape detection by immunity, ignore the normal checks on growth, alter its metabolic pathways for energy production, send signals that call other tissues to remodel the tumor's environment, and many other novel traits\autocite{weinberg07the-biology}.

\subsection*{Discovery and integration}

Innovation proceeds by the layering of new changes on top of the recent changes. Each particular change creates new context, favoring a new set of changes. Three evolutionary processes of cancer likely apply to many cases of evolutionary innovation.

First, an advantageous change enhances growth. Steady growth leads to the perception of abrupt origin. However, a single change by itself does not transform normal tissue into a cancer. Evolutionary innovation requires multiple changes. The early changes accumulate imperceptibly. 

Second, each change alters the context for future innovation. At some point, a single subsequent change could ignite growth. However, current evidence suggests a complex array of interacting changes that arise and spread over different timescales\autocite{frank07dynamics}. Advances in biological technology will eventually resolve the timing and the role of particular changes.

Third, as evolutionary change alters context, new pressures favor novel kinds of innovation. Sometimes, the novelty is itself a new generative mechanism that enhances the speed at which further novelty can be created. Or the novelty changes the way in which additional novelty integrates into the evolving population of cancerous cells.

The changing processes of discovery and integration in cancer likely arise in other evolutionary innovations. The following paragraphs describe a few examples for cancer. I then conclude by discussing aspects of language in relation to general properties of evolutionary innovation.

Suppose that an innovative trait would be favored, but it arises only one time per million cellular divisions. A tissue typically has far more than one million cells. So the trait arises many times in one round of cell division.  But only a few rare cells have the novel trait. 

The novel trait creates a context that would favor an additional innovation. Because only a few cells have the novel trait, it may take a very long time before the second innovation follows. However, if the initial trait spreads, then many cells would have the trait. The time before the second innovation would then be very short, because of the large size of the target population.

Rapid spread of the first trait may happen because it has a growth advantage and reproductively outcompetes other cells. Or the trait may spread if it produces a signal that transforms other cells to express the same trait. Much of cellular behavior arises by intercellular signalling. Transformation by novel signalling is a key aspect of evolutionary innovation in cancer progression\autocite{weinberg07the-biology}.

The discovery of a new trait is often discussed in terms of genetic mutation. Mutation couples two aspects in one stroke: the creation of novelty and the transmission of that novelty to future generations. However, one may have to wait a very long time for mutation to create a particular innovation. 

Alternatively, the novel trait may first appear by cellular adjustment to a novel environment\autocite{frank12nonheritable}. Initially, only a few cells may adjust to express the newly favored trait. Those cells gain an advantage, possibly transmitting to their descendants the tendency to adjust in the appropriate way. That process can favor rapid evolution of a novel trait that first appears by adjustment, or by learning, rather than by mutation\autocite{west-eberhard03developmental}. 

An environmental challenge may require two novel traits to arise simultaneously. For example, a novel cellular signal may require other cells to express a novel ability to respond to the signal. How do jointly synergistic traits evolve, if neither trait alone provides value\autocite{frank95the-origin,skyrms10signals:}? 

If some cells and their descendants remain spatially associated over time, then the group evolves almost like a single unit. The origin of the signal, initially by chance, strongly favors the recipient response. Signal and response may arise by one mutation then another. However, it may be a long time before two rare mutations arise.

Alternatively, different cells with the same genetics inevitably have a certain amount of randomness in the traits that they express. A population of cells that, by chance, expresses the right combination of novel signal and response traits will gain a growth advantage. 

Any genetic tendency to express the right trait combination will increase. Over time, the beneficial combination evolves to be expressed more frequently\autocite{frank11naturalb}. The process assimilates an initial tendency for random expression of traits into an increased genetic tendency to express the traits. Synergistic trait combinations can evolve relatively rapidly by this process when compared to the slow pace of origin by sequential mutations.

These ideas about innovation follow from classical evolutionary theory. We do not yet know exactly which aspects apply to particular cancers. However, technological advances will soon provide additional insight. 

\subsection*{Recursive hierarchy}

My discussion of evolutionary innovation and timescale for cancer applies broadly to any evolutionary system. Recursion unifies the conceptual frame.

First, natural selection recursively drives the spread of innovations. Given an input population, selection enhances the frequency of beneficial traits, producing a new output population. The output then becomes the input for another round of selection. An innovation with constant benefit increases by recursive multiplication, transforming constancy of benefit into explosive increase.

Second, an innovation can act by enhancing the rate at which additional new innovations are discovered. A discovery mechanism increases by selection when it associates with the beneficial innovations that it creates\autocite{otto02resolving}. Discovery applies recursively to each new generation.

Third, the trait of an individual may itself be a recursive system. Our bodies develop from the single-cell union of egg and sperm into approximately 30 trillion cells. Genetics does not specify the exact form of the adult. Instead, evolutionary history has built a developmental language applied recursively to the birth cell\autocite{carroll05endless}.

New evolutionary innovations arise by modification of the recursive developmental language. The encoding of traits in a recursive developmental language accelerates the discovery of innovations. 

In addition to the development of body form, other traits are also encoded by rules applied recursively. For example, our immune system combines recursive mechanisms to discover innovations and recursive mechanisms to select and enhance beneficial innovations\autocite{frank96the-design}. These recursive processes allow rapid discovery and expansion of novel defenses against infection.

\subsection*{Human language}

This hierarchy of recursive processes provides the framework for understanding evolutionary innovation. The \emph{origin} of human language falls naturally within this general evolutionary framework. However, the \emph{consequences} of human language add a new process of innovation. 

Before language, all evolutionary change had to follow a trajectory through the lineage of genes, a sufficiently stable molecular encoding of information to carry forward innovations. 

Human language created a parallel system to encode and transmit information. That parallel system follows the same general principles of recursion and innovation. However, the distinction between encoding by language or by molecules influences the recursive hierarchy and the consequences for innovation. The parallel systems of language and molecules interact, although the degree of coupling is controversial.

Language, as an innovation to the process of innovation, expands the recursive hierarchy and accelerates further innovation\autocite{cavalli-sforza81cultural,boyd85culture}. Evolutionary history has always been an evolving recursive hierarchy\autocite{frank96the-design}. When an evolutionary innovation alters the recursive hierarchy in a way that accelerates further innovation, then abrupt change often follows. 

Cancer, development and language differ. But they share the ways in which interacting recursive processes alter the timescale of innovation.

\subsection*{Competing interests}
No competing interests were disclosed.

\subsection*{Grant information}
National Science Foundation grant DEB--1251035 supports my research.

{\small\bibliographystyle{unsrtnat}
\bibliography{abrupt}}

\begin{thebibliography}{16}
\providecommand{\natexlab}[1]{#1}
\providecommand{\url}[1]{\texttt{#1}}
\expandafter\ifx\csname urlstyle\endcsname\relax
  \providecommand{\doi}[1]{doi: #1}\else
  \providecommand{\doi}{doi: \begingroup \urlstyle{rm}\Url}\fi

\bibitem[Smith and Harper(2013)]{smith13causes}
M.~P. Smith and D.~A.~T. Harper.
\newblock Causes of the {C}ambrian explosion.
\newblock \emph{Science}, 341:\penalty0 1355--1356, 2013.

\bibitem[Friedman(2009)]{friedman09the-meaning}
W.~E. Friedman.
\newblock The meaning of {D}arwin's `abominable mystery'.
\newblock \emph{American Journal of Botany}, 96:\penalty0 5--21, 2009.
\newblock \doi{10.3732/ajb.0800150}.

\bibitem[Tattersall(2016)]{tattersall16at-the-birth}
I.~Tattersall.
\newblock At the birth of language.
\newblock \emph{New York Review of Books}, {\rm August 18}:\penalty0 27--28,
  2016.

\bibitem[Berwick and Chomsky(2015)]{berwick15why-only}
R.~C. Berwick and N.~Chomsky.
\newblock \emph{Why Only Us: Language and Evolution}.
\newblock MIT Press, Boston, 2015.

\bibitem[Weinberg(2007)]{weinberg07the-biology}
R.~A. Weinberg.
\newblock \emph{The Biology of Cancer}.
\newblock Garland Science, New York, 2007.

\bibitem[Frank(2007)]{frank07dynamics}
S.~A. Frank.
\newblock \emph{Dynamics of {C}ancer: {I}ncidence, {I}nheritance, and
  {E}volution}.
\newblock Princeton University Press, Princeton, NJ, 2007.

\bibitem[Frank and Rosner(2012)]{frank12nonheritable}
S.~A. Frank and M.~R. Rosner.
\newblock Nonheritable cellular variability accelerates the evolutionary
  processes of cancer.
\newblock \emph{PLoS Biology}, 10:\penalty0 e1001296, 2012.

\bibitem[West-Eberhard(2003)]{west-eberhard03developmental}
M.~J. West-Eberhard.
\newblock \emph{{Developmental Plasticity and Evolution}}.
\newblock Oxford University Press, New York, 2003.

\bibitem[Frank(1995)]{frank95the-origin}
S.~A. Frank.
\newblock The origin of synergistic symbiosis.
\newblock \emph{Journal of Theoretical Biology}, 176:\penalty0 403--410, 1995.

\bibitem[Skyrms(2010)]{skyrms10signals:}
B.~Skyrms.
\newblock \emph{Signals: Evolution, Learning, and Information}.
\newblock Oxford University Press, New York, 2010.

\bibitem[Frank(2011)]{frank11naturalb}
S.~A. Frank.
\newblock Natural selection. {II}. {D}evelopmental variability and evolutionary
  rate.
\newblock \emph{Journal of Evolutionary Biology}, 24:\penalty0 2310--2320,
  2011.
\newblock \doi{10.1111/j.1420-9101.2011.02373.x}.

\bibitem[Otto and Lenormand(2002)]{otto02resolving}
S.~P. Otto and T.~Lenormand.
\newblock Resolving the paradox of sex and recombination.
\newblock \emph{Nature Reviews Genetics}, 3:\penalty0 252--261, 2002.

\bibitem[Carroll(2005)]{carroll05endless}
S.~B. Carroll.
\newblock \emph{Endless Forms Most Beautiful: The New Science of Evo Devo}.
\newblock W. W. Norton \& Company, Inc, New York, 2005.

\bibitem[Frank(1996)]{frank96the-design}
S.~A. Frank.
\newblock The design of natural and artificial adaptive systems.
\newblock In M.~R. Rose and G.~V. Lauder, editors, \emph{Adaptation}, pages
  451--505. Academic Press, San Diego, California, 1996.

\bibitem[Cavalli-Sforza and Feldman(1981)]{cavalli-sforza81cultural}
L.~L. Cavalli-Sforza and M.~W. Feldman.
\newblock \emph{Cultural {T}ransmission and {E}volution: {A} {Q}uantitative
  {A}pproach}.
\newblock Princeton University Press, Princeton, New Jersey, 1981.

\bibitem[Boyd and Richerson(1985)]{boyd85culture}
R.~Boyd and P.~J. Richerson.
\newblock \emph{Culture and the {E}volutionary {P}rocess}.
\newblock University of Chicago Press, Chicago, 1985.

\end{thebibliography}

\end{document}